# The PFDL-Model-Free Adaptive Predictive Control for a Class of Discrete-Time Nonlinear Systems

Feilong Zhang
State Key Laboratory of Robotics, Shenyang Institute of Automation, Chinese Academy of Sciences, Shenyang 110016, China

***Abstract*—In this paper, a novel partial form dynamic linearization (PFDL) data-driven model-free adaptive predictive control (MFAPC) method is proposed for a class of discrete-time single-input single-output nonlinear systems. It outperforms the current MFAC for that makes the prediction of the outputs of PFDL model over several steps and simultaneously uses more trajectory information in the future. The main contributions of this paper are that we combine the concept of MPC with MFAC together to propose a novel MFAPC method. Additionally, we prove the bounded-input bounded-output stability and tracking error convergence of the proposed method; Moreover, we have shown that the MFAPC will degenerate into MFAC as the prediction step *N*=1. The simulation and experiment are carried out to verify the effectiveness of the proposed MFAPC.**

## I. INTRODUCTION

Traditional feedback control methods and modern control theory methods have encountered many problems in practical applications. Most of them are typical model-based control methods and require the offline model of the systems in controller design [1]-[4]. However, the accurate physical model of the nonlinear time-varying system is hard to be identified in most industrial settings. Consequently, the idea of self-tuning control was firstly proposed by Kalman [5] in optimal control system design in 1958. Afterwards, minimum variance self-tuning regulator was proposed by Astorm and Wiittenmark, yet it is not applicable in non-minimum phase system for involving zero-pole cancellation [6], [7]. Thereafter, a generalized minimum variance controller was proposed by Clarke to extend the application in non-minimum phase system [8]. In addition, the stability and the convergence of several kinds of adaptive and generalized predictive control methods were analyzed by [9]-[15], which promotes a variety of adaptive control methods proposed and applied in industrial settings [16]-[19].

Nowadays, the data-driven model-free adaptive control (MFAC) has drawn much attention. Similar to above adaptive methods, it is not necessary to build the offline model of the system. The traditional ARMAX model is replaced by the equivalent dynamic linearization data models, which is shown as the increment form of the LTI DARMA model in [20]-[22]. The pseudo-gradient (PG) vector, whose components act as the coefficients of the equivalent dynamic linearization data models, is based on the deterministic estimation algorithms and merely estimated by the I/O measurement data of the controlled system [20]-[23]. Moreover, unmodeled dynamics do not exist in the data-driven model-free-adaptive control method, which gives a simplified discrete control structure to MFAC [22]. These advantages make it suitable for many practical applications through computer. For example, MFAC has been implemented in chemical industry, linear motor control and injection molding process, PH value control, and robotic welding process [20].

In order to further improve the behaviors and robustness of systems controlled by the current PFDL-MFAC method, we propose the PFDL-MFAPC, which can make full use of I/O measurement data in the past time to predict the output of the system and more future information of the reference trajectory to adjust the system input appropriately before the reference trajectory changes. The above advantages of the MFAPC can be attributed to that the index function of the MFAPC takes multiple prediction errors into consideration. On the contrary, the index function of the MFAC only lies in the current time. Besides, the MFAPC can be regarded as a matrix extension of the MFAC. The future coefficients of MFAPC need more iterations to predict, which can make further use of I/O measurement data in the past time and may improve the robustness of the system against the disturbance.

In control engineering community, the model predictive control (MPC) shows many superior properties and broad prospects in the robotic systems, such as, MIT's Cheetah 3 controlled by MPC can apply the right forces on the ground. However, MPC may not work well under model mismatches. Thereby, this motivates us to introduce the MFAPC by the combination of the MPC concept and MFAC together. More interestingly, this paper shows an important finding: the proposed PFDL-MFAPC can be considered as an elegant generalization of the current PFDL-MFAC with respect to the prediction step, sharing the identical structure, which hasn't been discussed so far, to the author's best knowledge. Along with this, PFDL-MFAPC has all the characteristics of the PFDL-MFAC, whose characteristics are detailed in [20]-[23].

The main contributions of this work are summarized as follows.

1) This paper proposes a method of PFDL-MFAPC with adjustable parameters and has shown that the proposed PFDL-MFAPC will degenerate into current PFDL-MFAC as *N*=1.

2) The bounded-input bounded-output (BIBO) stability and the monotonic convergence of the tracking error dynamics of the PFDL-MFAPC method are analyzed.

3) The effectiveness and merits of the proposed method are verified through the comparisons with the current MFAC in the simulation and experiment.

The rest of the paper is organized as follows. In Section II, the equivalent PFDL data predictive model is presented for a class of discrete time nonlinear systems. In Section III, we present the PFDL-MFAPC method design and its stability analysis results. In Section IV, the effectiveness of the proposed PFDL-MFAPC method are validated by the simulation and experiment. Section V gives the conclusions. At last, Appendix presents the detailed stability analysis of the proposed method.

## II. DYNAMIC LINEARIZATION DATA PREDICTIVE MODELS FOR DISCRETE-TIME NONLINEAR SYSTEMS

### A. System Model

In this section, an equivalent dynamic linearization data predictive model is given for general nonlinear discrete-time systems. Then, it is used in Sections III and IV to design and analyze the PFDL-MFAPC.

The discrete-time SISO nonlinear system is given as follows:

$$y(k+1)=f(y(k),\cdots,y(k-n_y),u(k),\cdots,u(k-n_u)) \quad (1)$$

where $f(\cdot)\in$ R is an unknown nonlinear function, $n_u$, $n_y\in$ Z represent the unknown orders of system input $u(k)$ and the system output $y(k)$ at time of $k$, respectively.

The PFDL of the nonlinear system (1) satisfies the following assumptions:

*Assumption 1:* The partial derivatives of $f(\cdots)$ with respect to control input $u(k)$, ⋯, $u(k-L)$ are continuous.

*Assumption 2:* System (1) satisfies the following generalized Lipschitz condition

$$|y(k_1+1)-y(k_2+1)|\le b\|\boldsymbol{U}(k_1)-\boldsymbol{U}(k_2)\| \quad (2)$$

where $\boldsymbol{U}(k)=[u(k),\cdots,u(k-L+1)]^T$ is a vector that contains control input within a time window $[k-L+1,k]$. $L$ $(1\le L\le n_u)$ is called pseudo orders of the system. For more detailed explanations about *Assumption 1* and *Assumption 2* please refer to [20], [21].

*Theorem 1:* For the non-linear system (1) satisfying *Assumptions 1* and *2*, there must exist a time-varying vector $\boldsymbol{\phi}_L(k)$ called PG vector; if $\Delta\boldsymbol{U}(k)\ne 0$, $1\le L$, system (1) can be transformed into the PFDL data model shown as follows

$$\Delta y(k+1)=\boldsymbol{\phi}_L^T(k)\Delta\boldsymbol{U}(k) \quad (3)$$

For any time $k$, we have $\|\boldsymbol{\phi}_L(k)\|\le b$, where $\boldsymbol{\phi}_L^T(k)=[\phi_1(k),\cdots,\phi_L(k)]$, $\Delta\boldsymbol{U}(k)=[\Delta u(k),\cdots,\Delta u(k-L+1)]^T$.

*Proof:* For details, please refer to [20], [21].

*Remark 1:* For detailed meaning and significances about the dynamic linearization data modeling method, please refer to [20][21].

The relationship between LTI DARMA model and the dynamic linearization data model is also presented in [20][21], which give the suggestions of how to choose the pseudo-orders $L$ of the model.

### B. Predictive System Model

Rewrite Equation (3) into the $N$ step forward prediction equation:

$$y(k+1)=y(k)+\boldsymbol{\phi}_L^T(k)\Delta\boldsymbol{U}(k) \quad (4)$$

Here, we define

$$\boldsymbol{A}=\begin{bmatrix}0 & & & \\ 1 & 0 & & \\ & \ddots & \ddots & \\ & & 1 & 0\end{bmatrix}_{L\times L} \qquad \boldsymbol{B}^T=\begin{bmatrix}1 & 0 & \cdots & 0\end{bmatrix}_{1\times L}$$

Based on (4), we have

$$\begin{aligned}
y(k+1)&=y(k)+\boldsymbol{\phi}_L^T(k)\Delta\boldsymbol{U}_L(k)\\
&=y(k)+\boldsymbol{\phi}_L^T(k)\boldsymbol{A}\Delta\boldsymbol{U}_L(k-1)+\boldsymbol{\phi}_L^T(k)\boldsymbol{B}\Delta u(k)\\
y(k+2)&=y(k+1)+\boldsymbol{\phi}_L^T(k+1)\Delta\boldsymbol{U}_L(k+1)\\
&=y(k)+\boldsymbol{\phi}_L^T(k)\boldsymbol{A}\Delta\boldsymbol{U}_L(k-1)\\
&\quad+\boldsymbol{\phi}_L^T(k)\boldsymbol{B}\Delta u(k)+\boldsymbol{\phi}_L^T(k+1)\boldsymbol{A}^2\Delta\boldsymbol{U}_L(k-1)\\
&\quad+\boldsymbol{\phi}_L^T(k+1)\boldsymbol{A}\boldsymbol{B}\Delta u(k)+\boldsymbol{\phi}_L^T(k+1)\boldsymbol{B}\Delta u(k+1)\\
&\;\;\vdots\\
y(k+N)&=y(k)+\sum_{i=0}^{N-1}\boldsymbol{\phi}_L^T(k+i)\boldsymbol{A}^{i+1}\Delta\boldsymbol{U}_L(k-1)\\
&\quad+\sum_{i=0}^{N-1}\boldsymbol{\phi}_L^T(k+i)\boldsymbol{A}^{i}\boldsymbol{B}\Delta u(k)\\
&\quad+\sum_{i=1}^{N-1}\boldsymbol{\phi}_L^T(k+i)\boldsymbol{A}^{i-1}\boldsymbol{B}\Delta u(k+1)\\
&\quad+\sum_{i=2}^{N-1}\boldsymbol{\phi}_L^T(k+i)\boldsymbol{A}^{i-2}\boldsymbol{B}\Delta u(k+2)\\
&\quad+\cdots+\sum_{i=N_u-1}^{N-1}\boldsymbol{\phi}_L^T(k+i)\boldsymbol{A}^{i-N_u+1}\boldsymbol{B}\Delta u(k+N_u-1)\\
&\quad+\cdots+\boldsymbol{\phi}_L^T(k+N-1)\boldsymbol{B}\Delta u(k+N-1)
\end{aligned} \quad (5)$$

where, $N$ is the predictive step length, $\Delta y(k+i)$ and $\Delta u(k+i)$ represent the increment values of the predictive output and the predictive input of the system in the future time $k+i$ ($i$=1,2,⋯, $N$ ), respectively. Here, we define $\boldsymbol{Y}_N(k)$, $\Delta\boldsymbol{Y}_N(k+1)$, $\Delta\boldsymbol{U}_N(k)$, $\Delta\boldsymbol{U}_{Nu}(k)$, $\bar{\boldsymbol{\Psi}}(k)$ and $\boldsymbol{\Psi}(k)$ as follows:

$$\Delta\boldsymbol{Y}_N(k+1)=\boldsymbol{Y}_N(k+1)-\boldsymbol{Y}_N(k)$$

$$\boldsymbol{Y}_N(k+1)=[y(k+1),\cdots,y(k+N)]_{1\times N}^T \qquad \boldsymbol{E}=[1,\cdots,1]_{1\times N}^T$$

$$\boldsymbol{\Psi}(k)_{N\times N}$$

$$=\begin{bmatrix}
\boldsymbol{\phi}_L^T(k)\boldsymbol{B} & 0 & \cdots & 0\\
\sum_{i=0}^{1}\boldsymbol{\phi}_L^T(k+i)\boldsymbol{A}^i\boldsymbol{B} & \boldsymbol{\phi}_L^T(k+1)\boldsymbol{B} & \vdots & \vdots\\
\vdots & \vdots & \ddots & \vdots\\
\sum_{i=0}^{N_u-1}\boldsymbol{\phi}_L^T(k+i)\boldsymbol{A}^i\boldsymbol{B} & \sum_{i=1}^{N_u-1}\boldsymbol{\phi}_L^T(k+i)\boldsymbol{A}^{i-1}\boldsymbol{B} & \ddots & 0\\
\vdots & \vdots & \ddots & \vdots\\
\sum_{i=0}^{N-1}\boldsymbol{\phi}_L^T(k+i)\boldsymbol{A}^i\boldsymbol{B} & \sum_{i=1}^{N-1}\boldsymbol{\phi}_L^T(k+i)\boldsymbol{A}^{i-1}\boldsymbol{B} & \cdots & \boldsymbol{\phi}_L^T(k+N-1)\boldsymbol{B}
\end{bmatrix}$$

$$\Delta \boldsymbol{U}_N(k)=[\Delta u(k),\cdots,\Delta u(k+N-1)]^T_{1\times N}$$

$$\Delta \boldsymbol{U}_{Nu}(k)=[\Delta u(k),\cdots,\Delta u(k+N_u-1)]^T_{1\times Nu}$$

$$\begin{aligned}\bar{\boldsymbol{\Psi}}(k)&=[\boldsymbol{A}\boldsymbol{\phi}_L(k),\sum_{i=0}^{1}\boldsymbol{A}^{i+1}\boldsymbol{\phi}_L(k+i)\boldsymbol{A}^{i+1},\cdots,\sum_{i=0}^{N_u-1}\boldsymbol{A}^{i+1}\boldsymbol{\phi}_L(k+i),\cdots,\\&\quad\sum_{i=0}^{N-1}\boldsymbol{A}^{i+1}\boldsymbol{\phi}_L(k+i)]^T\\&=[\bar{\boldsymbol{\Psi}}_1(k),\bar{\boldsymbol{\Psi}}_2(k),\cdots,\bar{\boldsymbol{\Psi}}_{L-1}(k),0]_{N\times L}\end{aligned}$$

where, $\bar{\boldsymbol{\Psi}}_j(k)$ is the $j$-th column of the $\bar{\boldsymbol{\Psi}}(k)$.

Then, (5) may be written as:

$$\boldsymbol{Y}_N(k+1)=\boldsymbol{E}y(k)+\boldsymbol{\Psi}(k)\Delta\boldsymbol{U}_N(k)+\bar{\boldsymbol{\Psi}}(k)\Delta\boldsymbol{U}_L(k-1) \quad (6)$$

$N_u$ is the control step length. If $\Delta u(k+j-1)=0$, $N_u<j\le N$, we can rewrite equation (6) into

$$\boldsymbol{Y}_N(k+1)=\boldsymbol{E}y(\mathrm{k})+\tilde{\boldsymbol{\Psi}}(k)\Delta\boldsymbol{U}_{N_u}(k)+\bar{\boldsymbol{\Psi}}(k)\Delta\boldsymbol{U}_L(k-1) \quad (7)$$

Where

$$\tilde{\boldsymbol{\Psi}}(k)_{N\times Nu}=\begin{bmatrix}\boldsymbol{\phi}_L^T(k)\boldsymbol{B} & & & \\ \sum_{i=0}^{1}\boldsymbol{\phi}_L^T(k+i)\boldsymbol{A}^i\boldsymbol{B} & \boldsymbol{\phi}_L^T(k+1)\boldsymbol{B} & \vdots & \\ \vdots & \vdots & \vdots & \vdots\\ \sum_{i=0}^{Nu-1}\boldsymbol{\phi}_L^T(k+i)\boldsymbol{A}^i\boldsymbol{B} & \sum_{i=1}^{Nu-1}\boldsymbol{\phi}_L^T(k+i)\boldsymbol{A}^{i-1}\boldsymbol{B} & \cdots & \boldsymbol{\phi}_L^T(k+N_u-1)\\ \vdots & \vdots & \vdots & \vdots\\ \sum_{i=0}^{N-1}\boldsymbol{\phi}_L^T(k+i)\boldsymbol{A}^i\boldsymbol{B} & \sum_{i=1}^{N-1}\boldsymbol{\phi}_L^T(k+i)\boldsymbol{A}^{i-1}\boldsymbol{B} & & \sum_{i=Nu-1}^{N-1}\boldsymbol{\phi}_L^T(k+i)\boldsymbol{A}^{i-Nu+1}\boldsymbol{B}\end{bmatrix}$$

## III. MODEL-FREE ADAPTIVE PREDICTIVE CONTROL DESIGN AND STABILITY ANALYSIS

In this section, the design of PFDL-MFAPC method will firstly be presented. In addition, the relationship between the PFDL-MFAPC and PFDL-MFAC is presented. After that, the stability analysis with some necessary Theorems and Lemma are presented.

### A. *Design of PFDL Model Free Adaptive Predictive Control*

A weighted control input index function is given as

$$\begin{aligned}J&=\left[\boldsymbol{Y}_N^*(k+1)-\boldsymbol{Y}_N(k+1)\right]^T\left[\boldsymbol{Y}_N^*(k+1)-\boldsymbol{Y}_N(k+1)\right]\\&\quad+\boldsymbol{\lambda}\Delta\boldsymbol{U}_{Nu}^T(k)\Delta\boldsymbol{U}_{Nu}(k)\end{aligned} \quad (8)$$

where, λ is the weighted constant; $\boldsymbol{Y}_N^*(k+1)=\left[y^*(k+1),\cdots,y^*(k+N)\right]^T$ is the desired system output vector, where $y^*(k+i)$ is the desired output of the system at the future time of $(k+i)$ ($i$=1,2,···, $N$).

Considering that the PG vector can be obtained, combining Equation (7) with Equation (8) and solving the optimization condition $\partial J/\partial\Delta\boldsymbol{U}_{Nu}(k)=0$, we have the optimal output solution:

$$\begin{aligned}\Delta\boldsymbol{U}_{Nu}^0(k)&=\left[\tilde{\boldsymbol{\Psi}}(k)^T\tilde{\boldsymbol{\Psi}}(k)+\lambda\boldsymbol{I}\right]^{-1}\tilde{\boldsymbol{\Psi}}^T(k)\bullet\\&\quad\left[\rho_1(\boldsymbol{Y}_N^*(k+1)-\boldsymbol{E}(k)y(k))-\bar{\boldsymbol{\Psi}}(k)\boldsymbol{\Lambda}\Delta\boldsymbol{U}_L(k-1)\right]\end{aligned} \quad (9)$$

where, $\rho_i<1$ $(i=1,2,\cdots,L+1)$ are the adjustable parameters, $\rho_1$ and $\boldsymbol{\Lambda}=diag[\rho_2,\cdots,\rho_{L+1}]$ are introduced to make the controller algorithm more flexible and to analysis the stability of the system. It also helps us to change the behavior of the system as the performance is not acceptable. Then, we have the optimal current input

$$u^0(k)=u^0(k-1)+\boldsymbol{g}^T\Delta\boldsymbol{U}_{Nu}^0(k) \quad (10)$$

where $\boldsymbol{g}=\left[1,0,\cdots,0\right]^T$.

*Remark 2:* The unknown $\boldsymbol{\phi}_L(k+i)$ ( $i$= 0, 1, 2, ···, $N$-1), which make up unknown $\tilde{\boldsymbol{\Psi}}_{Nu}(k)$ and $\bar{\boldsymbol{\Psi}}(k)$ in Equation (9), need to be replaced by their estimated and predicted values $\hat{\boldsymbol{\phi}}_L(k+i)$. [21], [23] give the projection algorithm to estimate $\boldsymbol{\phi}_L(k)$ and reset the $\hat{\boldsymbol{\phi}}_L(k)$ by the initial vector according to the following algorithm.

$$\hat{\boldsymbol{\phi}}_L(k)=\hat{\boldsymbol{\phi}}_L(1)\ ,\text{ if }\left|\hat{\boldsymbol{\phi}}_L(k)\right|\le\boldsymbol{\varepsilon}\text{ or }sign(\hat{\phi}_1(k))\ne sign(\hat{\phi}_1(1)) \quad (11)$$

[21], [23] give the $\hat{\boldsymbol{\phi}}_L(k+i)$, $i$= 1, 2, ···, $N$-1 by the data-driven multi-level hierarchical forecasting method. From the above references, we know that the $\hat{\boldsymbol{\phi}}_L(k+i)$ ($i$= 0, 1, 2, ···, $N$-1), which are the linear combination of the $\hat{\boldsymbol{\phi}}_L(k)$, $\hat{\boldsymbol{\phi}}_L(k-1)$, ···, $\hat{\boldsymbol{\phi}}_L(k-n_p+1)$, are bounded. Let us define $\hat{\tilde{\boldsymbol{\Psi}}}(k)$ and $\hat{\bar{\boldsymbol{\Psi}}}(k)$ as the estimated matrixes of the $\tilde{\boldsymbol{\Psi}}(k)$ and $\bar{\boldsymbol{\Psi}}(k)$, respectively. Then, according to the definition of the norms of matrix, the norms of $\hat{\tilde{\boldsymbol{\Psi}}}(k)$ and $\hat{\bar{\boldsymbol{\Psi}}}(k)$ are bounded.

Then we obtain the proposed PFDL-MFAPC control law

$$\begin{aligned}\Delta\boldsymbol{U}_{Nu}(k)&=\left[\hat{\tilde{\boldsymbol{\Psi}}}(k)^T\hat{\tilde{\boldsymbol{\Psi}}}(k)+\lambda\boldsymbol{I}\right]^{-1}\hat{\tilde{\boldsymbol{\Psi}}}^T(k)\bullet\\&\quad\left[\rho_1(\boldsymbol{Y}_N^*(k+1)-\boldsymbol{E}(k)y(k))-\hat{\bar{\boldsymbol{\Psi}}}(k)\boldsymbol{\Lambda}\Delta\boldsymbol{U}_L(k-1)\right]\end{aligned} \quad (12)$$

The actual control law at the current instant is

$$u(k)=u(k-1)+\boldsymbol{g}^T\Delta\boldsymbol{U}_{Nu}(k) \quad (13)$$

*Remark 3:* The methods of how to choose $N$ and $N_u$ are detailed in [21].

*Remark 4:* The special cases of the proposed PFDL-MFAPC method are shown below.

When $N_u=1$, we have the following simplified control output, which does not have the inverse calculation of matrix

$$\begin{aligned}\Delta\boldsymbol{U}_{Nu}(k)&=\frac{1}{[\hat{\tilde{\boldsymbol{\Psi}}}^T(k)]_{1\times N}[\hat{\tilde{\boldsymbol{\Psi}}}(k)]_{N\times 1}+\lambda}[\hat{\tilde{\boldsymbol{\Psi}}}^T(k)]_{1\times N}\bullet\\&\quad[\rho_1(\boldsymbol{Y}_N^*(k+1)-\boldsymbol{E}y(k))-\\&\quad[\bar{\boldsymbol{\Psi}}(k)]_{N\times L}\begin{bmatrix}\rho_2 & & \\ & \ddots & \\ & & \rho_{L+1}\end{bmatrix}\begin{bmatrix}\Delta u(k-1)\\ \vdots\\ \Delta u(k-L+1)\end{bmatrix}]\end{aligned} \quad (14)$$

When $N=1$ and the corresponding $N_u=1$, the PFDL-MFAPC degenerates into the PFDL-MFAC.

$$\Delta u(k)=\frac{\hat{\phi}_1(k)}{\lambda+|\phi_1(k)|^2}[\rho_1(y^*-y(k)) -[\hat{\phi}_2(k),\cdots,\hat{\phi}_L(k),0]\begin{bmatrix}\rho_2 & & \\ & \ddots & \\ & & \rho_{L+1}\end{bmatrix}\begin{bmatrix}\Delta u(k-1)\\ \vdots \\ \Delta u(k-L+1)\end{bmatrix}]$$
$$=\frac{\rho_1\hat{\phi}_1(k)(y^*-y(k))}{\lambda+|\phi_1(k)|^2}-\frac{\hat{\phi}_1(k)\sum_{i=2}^{L}\rho_i\hat{\phi}_i(k)\Delta u(k-i+1)}{\lambda+|\phi_1(k)|^2} \tag{15}$$

From (15), we can conclude that the proposed PFDL-MFAPC can be considered as an elegant extension of the current PFDL-MFAC, whose meaning and analysis are shown in [20]-[22].

## B. *Stability Analysis of MFAPC*

This section gives some Lemmas, assumptions, and the proof of stability of PFDL-MFAPC.

*Lemma 1*:[24] Let $\boldsymbol{A}=\begin{bmatrix} a_1 & \cdots & \cdots & a_{L-1} & a_L \\ 1 & 0 & \cdots & 0 & 0 \\ 0 & 1 & \cdots & 0 & 0 \\ \vdots & \vdots & \vdots & \vdots & \vdots \\ 0 & 0 & \cdots & 1 & 0 \end{bmatrix}$. If $\sum_{i=1}^{L}|a_i|<1$, then $\sigma(\boldsymbol{A})<1$, where $\sigma(\boldsymbol{A})$ is the spectral radius of $\boldsymbol{A}$.

*Lemma 2*: When $\boldsymbol{A}\in R^{n\times n}$, for any given $\varepsilon>0$, there exists an induced consistent matrix norm such that

$$\|\boldsymbol{A}\|_v \le \sigma(\boldsymbol{A})+\varepsilon$$

where $\sigma(\boldsymbol{A})$ is the spectral radius of $\boldsymbol{A}$.

*Theorem 2*: If the system is described by (1) and controlled by the MFAPC method (9)-(10) with the desired trajectory $y_d(k)=y_d=$ constant, there exists a $\lambda_{\min}$, such that when $\lambda>\lambda_{\min}$, it guarantees: 1) $\lim_{k\to\infty}|y(k+1)-y^*|=0$; 2) the control system is BIBO stability.

*Proof*: Appendix presents the proof of *Theorem 2*, which is inspired by [20][21].

## IV. SIMULATIONS

[21], [25] give a number of examples to compare MFAC with other typical DDC methods, like data-driven PID (DD-PID), iterative feedback tuning (IFT), and virtual reference feedback tuning (VRFT). The conclusion is that the tracking performance of MFAC is better than the above methods in its simulations. Therefore, we only need to show the effectiveness and the advantages of MFAPC methods by the comparison with MFAC.

Example 1: We choose an example from [21] to make comparisons between MFAPC and MFAC, and the following discrete-time SISO nonlinear structure-varying system is considered.

$$y(k)=\begin{cases}\dfrac{2.5y(k-1)y(k-2)}{1+y^2(k-1)+y^2(k-2)}+1.2u(k-1)+1.4u(k-2) \\ \quad +0.7\sin(0.5(y(k-1)+y(k-2))) \qquad 0<k\le 200 \\ -0.1y(k-1)-0.2y(k-2)-0.3y(k-3)+0.1u(k-1) \\ \quad +0.02\,\mathrm{u}(k-1)+0.03\,\mathrm{u}(k-1) \qquad 200<k\le 400\end{cases} \tag{16}$$

The system is structure-varying and discontinuous, and we suppose that the system is unknown to the controller design process. The desired output trajectory is

$$y^*(k+1)=5\times(-1)^{round(k/80)}, 1\le k\le 400$$

The controller parameters and initial setting for both the PFDL-MFAPC and PFDL-MFAC are listed in Table I, and all of them should be the same with [21].

We make comparisons among PFDL-MFAPC, PFDL-MFAC and the PID. [21] gives an appropriate group of PID parameters: $k_P$=0.15, $T_I$=0.5, $T_D$=0. The comparisons of tracking performance are shown in Fig. 1. The control inputs of these methods are shown in Fig. 2. The components of the PG estimation of both methods are shown in Fig. 3. The performance indexes for MFAPC and MFAC are shown in TABLE II.

TABLE I Parameter Settings for PFDL-MFAC and PFDL-MFAPC

| Parameter | MFAC | MFAPC |
|---|---|---|
| Order | L=3 | L=2 |
| $\lambda$, $\eta$, $\mu$ | 0.01, 0.5, 2 | 0.01, 0.5, 2 |
| $\rho_{1,2,3}$ | [0.5, 0.5, 0.5] | [0.5, 0.5, 0.5] |
| Initial value $\hat{\phi}_L(1)$ | [1, 0, 0] | [1, 0] |
| $u(0:5)$ | $(0,0,0,0,0)$ | $(0,0,0,0,0)$ |
| $y(0:5)$ | $(0,0,0,1,0)$ | $(0,0,0,1,0)$ |
| Predictive step N | 1 (No choice) | 3 |
| Control step $N_u$ | 1 (No choice) | 3 |

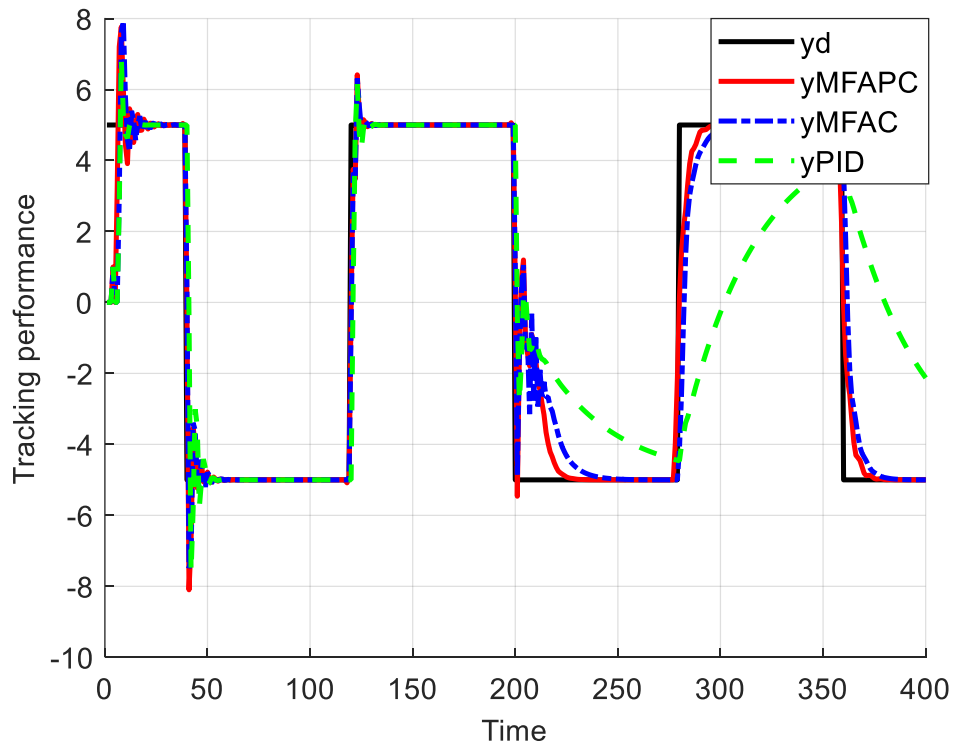

Figure. 1 Tracking performance

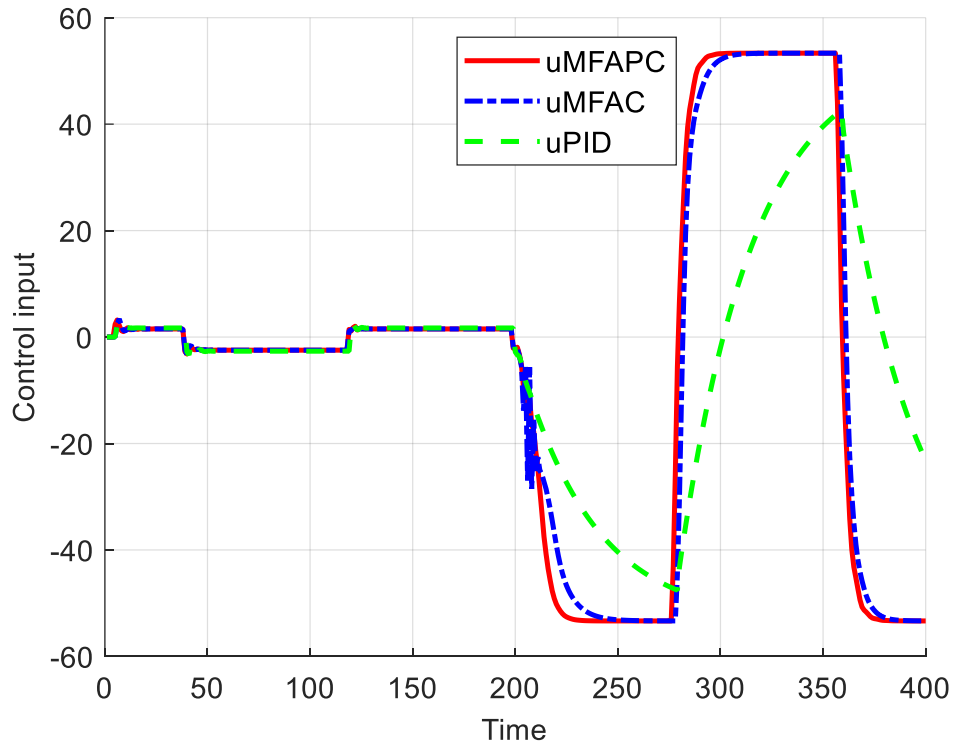

Fig. 2 Control input

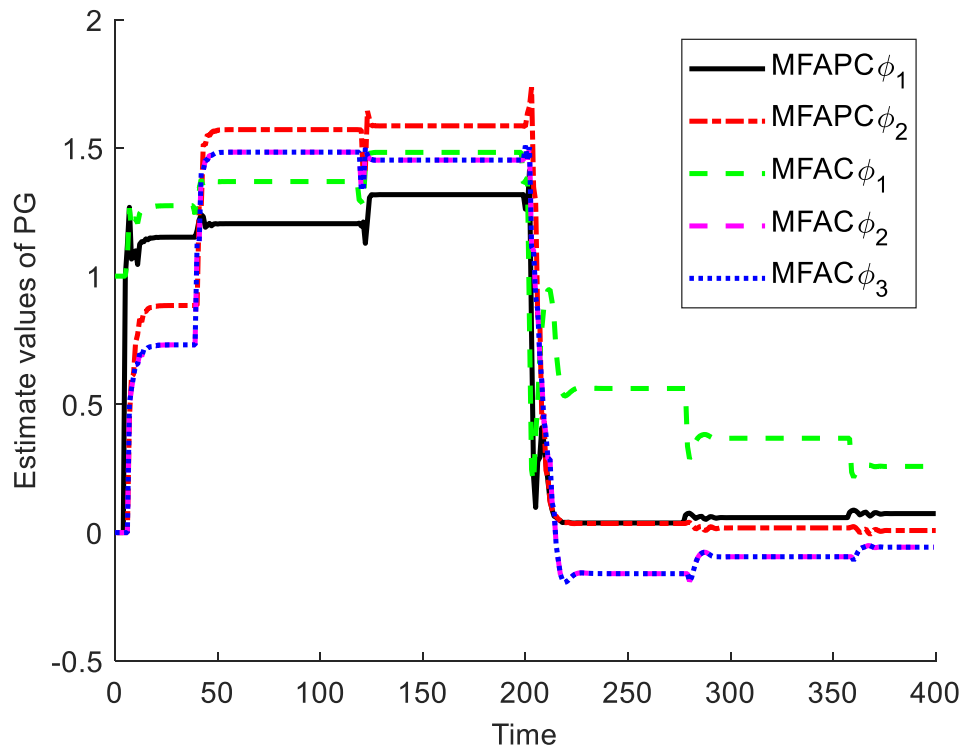

Fig. 3 Estimated value of PG

TABLE II Performance Indexes for MFAPC and MFAC

| | MFAC | MFAPC |
|---|---|---|
| $eITAE=\sum_{k=1}^{N}\lvert e(k)\rvert$ | 214.6279 | 165.1620 |

From Fig. 1 and TABLE II, we can see that the respond speed and the precision of the systems controlled by MFAPC is better than that controlled by MFAC, and the systems controlled by PID cannot converge well after the time 200. The above advantages can be attributed to that PFDL-MFAPC can make full use of I/O measurement data in the past time and use more future information of the reference trajectory.

## V. CONCLUSION

A novel model-free adaptive predictive control (MFAPC) method with adjustable parameters is proposed for a class of discrete-time single-input and single-output nonlinear systems. Then, we show the relationship between the PFDL-MFAC and the proposed PFDL-MFAPC. The bounded-input bounded-output (BIBO) stability analysis and the tracking error monotonic convergence of the MFAPC method are analyzed by the contraction mapping technique. The effectiveness of the proposed method has been illustrated by the simulation and experiment.

## APPENDIX: PROOF OF THEOREM 2

This section proves the convergence of the tracking error and the BIBO stability of the system controlled by the proposed PFDL-MFAPC.

We first define $\boldsymbol{P}=\boldsymbol{g}^T\left[\hat{\hat{\boldsymbol{\Psi}}}^T(k)\hat{\hat{\boldsymbol{\Psi}}}(k)+\lambda\boldsymbol{I}\right]^{-1}\hat{\hat{\boldsymbol{\Psi}}}^T(k)$.

According to Section II, we can express $\hat{\hat{\boldsymbol{\Psi}}}(k)$ as

$\hat{\hat{\boldsymbol{\Psi}}}(k)=[\hat{\hat{\boldsymbol{\Psi}}}_1(k),\hat{\hat{\boldsymbol{\Psi}}}_2(k),\cdots,\hat{\hat{\boldsymbol{\Psi}}}_{L-1}(k),0]_{N\times L}$,

From (12) and (13), we have

$$
\begin{aligned}
\Delta\boldsymbol{U}_L(k)&=[\Delta u(k),\cdots,\Delta u(k-L+1)]^T\\
&=\boldsymbol{A}(k)[u(k-1),\cdots,u(k-L)]^T\\
&\quad+\rho_1\boldsymbol{g}^T\left[\hat{\hat{\boldsymbol{\Psi}}}^T(k)\hat{\hat{\boldsymbol{\Psi}}}(k)+\lambda\boldsymbol{I}\right]^{-1}\hat{\hat{\boldsymbol{\Psi}}}^T(k)\boldsymbol{EC}e(k)\\
&=\boldsymbol{A}(k)\Delta\boldsymbol{U}_L(k-1)+\rho_1\boldsymbol{PEC}e(k)
\end{aligned}
\tag{17}
$$

Then, $\boldsymbol{A}(k)$ may be rewritten as

$$
\boldsymbol{A}(k)=\begin{bmatrix}-\boldsymbol{g}^T\left[\hat{\hat{\boldsymbol{\Psi}}}^T(k)\hat{\hat{\boldsymbol{\Psi}}}(k)+\lambda\boldsymbol{I}\right]^{-1}\hat{\hat{\boldsymbol{\Psi}}}^T(k)\hat{\hat{\boldsymbol{\Psi}}}(k)\boldsymbol{\Lambda}\\ \boldsymbol{I}_{(L-1)\times(L-1)}\ \vdots\ \boldsymbol{0}_{(L-1)\times 1}\end{bmatrix}_{L\times L}
=\begin{bmatrix}-\rho_2\boldsymbol{P}\hat{\hat{\boldsymbol{\Psi}}}_1(k) & -\rho_3\boldsymbol{P}\hat{\hat{\boldsymbol{\Psi}}}_2(k) & \cdots & -\rho_L\boldsymbol{P}\hat{\hat{\boldsymbol{\Psi}}}_{L-1}(k) & 0\\ 1 & 0 & \cdots & 0 & 0\\ 0 & 1 & \cdots & 0 & 0\\ \vdots & \vdots & \vdots & \vdots & \vdots\\ 0 & 0 & \cdots & 1 & 0\end{bmatrix}_{L\times L}
\tag{18}
$$

$$\boldsymbol{C}=[1,0,\cdots,0]_{L\times 1}^T$$

According to the sum of the first row of $\boldsymbol{A}(k)$ and the matrix norm inequalities between $\|\bullet\|_\infty$ and $\|\bullet\|_2$, we have

$$
\begin{aligned}
&\sum_{i=2}^{L}\left|\rho_i\boldsymbol{P}\hat{\hat{\boldsymbol{\Psi}}}_i(k)\right|\le(\max_{i=2,\cdots,L+1}\rho_i)\sum_{i=1}^{L-1}\left|\boldsymbol{P}\hat{\hat{\boldsymbol{\Psi}}}_i(k)\right|\\
&\le(\max_{i=2,\cdots,L+1}\rho_i)\left\|\left[\hat{\hat{\boldsymbol{\Psi}}}(k)^T\hat{\hat{\boldsymbol{\Psi}}}(k)+\lambda\boldsymbol{I}\right]^{-1}\hat{\hat{\boldsymbol{\Psi}}}^T(k)\hat{\hat{\boldsymbol{\Psi}}}(k)\right\|_\infty\\
&\le(\max_{i=2,\cdots,L+1}\rho_i)\sqrt{N_u}\left\|\left[\hat{\hat{\boldsymbol{\Psi}}}(k)^T\hat{\hat{\boldsymbol{\Psi}}}(k)+\lambda\boldsymbol{I}\right]^{-1}\right\|_2\left\|\hat{\hat{\boldsymbol{\Psi}}}^T(k)\right\|_\infty\left\|\hat{\hat{\boldsymbol{\Psi}}}(k)\right\|_\infty
\end{aligned}
\tag{19}
$$

where, $\hat{\hat{\boldsymbol{\Psi}}}^T(k)\hat{\hat{\boldsymbol{\Psi}}}(k)$ is a symmetric semi-positive matrix, meaning that $\left[\hat{\hat{\boldsymbol{\Psi}}}^T(k)\hat{\hat{\boldsymbol{\Psi}}}(k)+\lambda\boldsymbol{I}\right]$ is a symmetric positive matrix, so we have $\left[\left(\hat{\hat{\boldsymbol{\Psi}}}^T(k)\hat{\hat{\boldsymbol{\Psi}}}(k)+\lambda\boldsymbol{I}\right)^{-1}\right]^T=\left(\hat{\hat{\boldsymbol{\Psi}}}^T(k)\hat{\hat{\boldsymbol{\Psi}}}(k)+\lambda\boldsymbol{I}\right)^{-1}$; $\|\bullet\|_\infty$ is the maximum row sum matrix norm (row-sum norm). $\|\bullet\|_2$ is the spectral norm of matrix.

Assume the eigenvalues of $\hat{\hat{\boldsymbol{\Psi}}}^T(k)\hat{\hat{\boldsymbol{\Psi}}}(k)$ are $b_i\ge 0$, $i=1,\cdots,N_u$, so the eigenvalues of $\left[\hat{\hat{\boldsymbol{\Psi}}}^T(k)\hat{\hat{\boldsymbol{\Psi}}}(k)+\lambda\boldsymbol{I}\right]$ are $\lambda+b_i>0$, $i=1,\cdots,N_u$, then the eigenvalues of $[\hat{\hat{\boldsymbol{\Psi}}}^T(k)\hat{\hat{\boldsymbol{\Psi}}}(k)+\lambda\boldsymbol{I}]^{-1}$ are $\frac{1}{\lambda+b_i}>0$, $i=1,\cdots,N_u$. Therefore, we have

$$
\begin{aligned}
&\left\|\left[\hat{\hat{\boldsymbol{\Psi}}}(k)^T\hat{\hat{\boldsymbol{\Psi}}}(k)+\lambda\boldsymbol{I}\right]^{-1}\right\|_2\\
&=\sqrt{\sigma\left(\left(\left[\hat{\hat{\boldsymbol{\Psi}}}(k)^T\hat{\hat{\boldsymbol{\Psi}}}(k)+\lambda\boldsymbol{I}\right]^{-1}\right)^T\left[\hat{\hat{\boldsymbol{\Psi}}}(k)^T\hat{\hat{\boldsymbol{\Psi}}}(k)+\lambda\boldsymbol{I}\right]^{-1}\right)}\\
&=\sqrt{\sigma\left(\left(\left[\hat{\hat{\boldsymbol{\Psi}}}(k)^T\hat{\hat{\boldsymbol{\Psi}}}(k)+\lambda\boldsymbol{I}\right]^{-1}\right)^2\right)}=\frac{1}{\min_{i=1,\cdots Nu}\{\lambda+b_i\}}
\end{aligned}
\tag{20}
$$

Then, we can get (21) by combining (19) and (20).

$$\sum_{i=2}^{L}\left|\boldsymbol{P}\hat{\hat{\boldsymbol{\Psi}}}_i(k)\right| \le \sqrt{N_u}\left\|\left[\hat{\hat{\boldsymbol{\Psi}}}(k)^T\hat{\hat{\boldsymbol{\Psi}}}(k)+\lambda\boldsymbol{I}\right]^{-1}\right\|_2\left\|\hat{\hat{\boldsymbol{\Psi}}}^T(k)\right\|_\infty\left\|\hat{\hat{\boldsymbol{\Psi}}}(k)\right\|_\infty$$
$$\le \sqrt{N_u}\,\frac{1}{\min\limits_{i=1,\cdots Nu}\{\lambda+b_i\}}\left\|\hat{\hat{\boldsymbol{\Psi}}}^T(k)\right\|_\infty\left\|\hat{\hat{\boldsymbol{\Psi}}}(k)\right\|_\infty \tag{21}$$

Assume that the maximum row sum matrix norm of $\hat{\hat{\boldsymbol{\Psi}}}^T(k)$ can be got at the $s$-th row, then we can obtain that $\left\|\hat{\hat{\boldsymbol{\Psi}}}^T(k)\right\|_\infty = \sum_{j=1}^{N}\sum_{i=s-1}^{j-1}\boldsymbol{\phi}_L^T(k+i)\boldsymbol{A}^{i-s+1}\boldsymbol{B}$ is bounded. Suppose that the maximum row sum matrix norm of $\hat{\hat{\boldsymbol{\Psi}}}(k)$ can be got at the $s_1$-th row, then it is clear that $\left\|\hat{\hat{\boldsymbol{\Psi}}}(k)\right\|_\infty = \sum_{i=0}^{s_1-1}\boldsymbol{\phi}_L^T(k+i)\boldsymbol{A}^{i+1}$ is bounded. Therefore, there exists a positive $\lambda_{\min 1}$, if $\lambda > \lambda_{\min 1}$, we can obtain the following inequation:

$$\left[\sum_{i=1}^{L-1}\left|\boldsymbol{P}\hat{\hat{\boldsymbol{\Psi}}}_i(k)\right|\right]^{\frac{1}{L-1}} \le \left[\sqrt{N_u}\,\frac{1}{\min\limits_{i=1,\cdots Nu}\{\boldsymbol{\lambda}+b_i\}}\left\|\hat{\hat{\boldsymbol{\Psi}}}^T(k)\right\|_\infty\left\|\hat{\hat{\boldsymbol{\Psi}}}(k)\right\|_\infty\right]^{\frac{1}{L-1}}$$
$$\le M_4 < 1 \tag{22}$$

Given $0<\rho_2<1$, $\cdots$, $0<\rho_{L+1}<1$, we have $(\max\limits_{i=2,\cdots,L+1}\rho_i)<1$. Hence, we have

$$\sum_{i=1}^{L-1}\rho_{i+1}\left|\boldsymbol{P}\hat{\hat{\boldsymbol{\Psi}}}_i(k)\right| \le (\max_{i=2,\cdots,L+1}\rho_i)\sum_{i=1}^{L-1}\left|\boldsymbol{P}\hat{\hat{\boldsymbol{\Psi}}}_i(k)\right|$$
$$\le (\max_{i=2,\cdots,L+1}\rho_i)M_4^{L-1} < 1 \tag{23}$$

According to Lemma 1 and (24), we can see that the sum of the absolute values of each element in the first row of matrix $\boldsymbol{A}(k)$ is less than 1. Then, it is obvious that all the eigenvalues of $\boldsymbol{A}(k)$ satisfy $|z|<1$.The characteristic equation of $\boldsymbol{A}(k)$ is

$$z^L + \rho_2\mathbf{P}\hat{\hat{\boldsymbol{\Psi}}}_1(k)z^{L-1} + \cdots + \rho_L\mathbf{P}\hat{\hat{\boldsymbol{\Psi}}}_{L-1}(k)z = 0 \tag{24}$$

Based on $|z|<1$ and (25), we have the following inequation:

$$|z|^{L-1} \le \sum_{i=1}^{L-1}\rho_{i+1}\left|\boldsymbol{P}\hat{\hat{\boldsymbol{\Psi}}}_i(k)\right||z|^{L-i}$$
$$\le \sum_{i=1}^{L-1}\rho_{i+1}\left|\boldsymbol{P}\hat{\hat{\boldsymbol{\Psi}}}_i(k)\right| \le (\max_{i=2,\cdots,L+1}\rho_i)M_4^{L-1} < 1 \tag{25}$$

which means $|z| \le (\max\limits_{i=2,\cdots,L+1}\rho_i)^{1/L-1}M_1 < 1$. Hence, according to Lemma 2 and (25), there exists an arbitrarily small positive ε that makes the following inequation hold.

$$\|\boldsymbol{A}(k)\|_v \le \sigma(\boldsymbol{A}(k)) + \varepsilon \le (\max_{i=2,\cdots,L+1}\rho_i)^{1/L+1}M_1 + \varepsilon < 1 \tag{26}$$

where $\|\boldsymbol{A}(k)\|_v$ is the compatible norm of $\boldsymbol{A}(k)$. Let $d_2 = (\max\limits_{i=2,\cdots,L+1}\rho_i)^{1/L-1}M_1$.

Here, $\boldsymbol{PE} = \boldsymbol{g}^T\left[\hat{\hat{\boldsymbol{\Psi}}}(k)^T\hat{\hat{\boldsymbol{\Psi}}}(k)+\lambda\boldsymbol{I}\right]^{-1}\hat{\hat{\boldsymbol{\Psi}}}^T(k)\boldsymbol{E}$ is a number that equals to the sum of the each elements in the first row of $\left[\hat{\hat{\boldsymbol{\Psi}}}^T(k)\hat{\hat{\boldsymbol{\Psi}}}(k)+\lambda\boldsymbol{I}\right]^{-1}\hat{\hat{\boldsymbol{\Psi}}}^T(k)$. Then, we have

$$\boldsymbol{g}^T\left[\hat{\hat{\boldsymbol{\Psi}}}^T(k)\hat{\hat{\boldsymbol{\Psi}}}(k)+\lambda\boldsymbol{I}\right]^{-1}\hat{\hat{\boldsymbol{\Psi}}}^T(k)\boldsymbol{E}$$
$$\le \left\|\left[\hat{\hat{\boldsymbol{\Psi}}}^T(k)\hat{\hat{\boldsymbol{\Psi}}}(k)+\lambda\boldsymbol{I}\right]^{-1}\hat{\hat{\boldsymbol{\Psi}}}^T(k)\right\|_\infty$$
$$\le \sqrt{N_u}\left\|\left[\hat{\hat{\boldsymbol{\Psi}}}(k)^T\hat{\hat{\boldsymbol{\Psi}}}(k)+\lambda\boldsymbol{I}\right]^{-1}\right\|_2\left\|\hat{\hat{\boldsymbol{\Psi}}}^T(k)\right\|_\infty \tag{27}$$
$$= \sqrt{N_u}\,\frac{1}{\min\limits_{i=1,\cdots Nu}\{\lambda+b_i\}}\left\|\hat{\hat{\boldsymbol{\Psi}}}^T(k)\right\|_\infty$$

Similar to the proof process of (22), there exists positive $\lambda_{\min 2}$ and $M_2$, such that $\lambda > \lambda_{\min 2}$, then we obtain the following two inequations

$$0 < M_1 \le \boldsymbol{g}^T\left[\hat{\hat{\boldsymbol{\Psi}}}^T(k)\hat{\hat{\boldsymbol{\Psi}}}(k)+\lambda\boldsymbol{I}\right]^{-1}\hat{\hat{\boldsymbol{\Psi}}}^T(k)\boldsymbol{E} \le M_2 < 1 \tag{28}$$

$$d_3 \triangleq \rho_1 M_2\left\|\boldsymbol{\phi}_L^T(i+1)\right\|_v < 0.5 \tag{29}$$

Because of $|\Delta\boldsymbol{U}_L(0)|_v = 0$, taking the norm of (18) and combining (26) and (28), we have

$$|\Delta\boldsymbol{U}_L(k)|_v = |\boldsymbol{A}(k)|_v\,|\Delta\boldsymbol{U}_L(k-1)|_v$$
$$+\left|\rho_1\boldsymbol{g}^T\left[\hat{\hat{\boldsymbol{\Psi}}}(k)^T\hat{\hat{\boldsymbol{\Psi}}}(k)+\lambda\boldsymbol{I}\right]^{-1}\hat{\hat{\boldsymbol{\Psi}}}^T(k)\boldsymbol{E}\right||e(k)|$$
$$= d_2\,|\Delta\boldsymbol{U}_L(k-1)|_v + \rho_1 M_2\,|e(k)| \tag{30}$$
$$\vdots$$
$$= \rho_1 M_2\sum_{i=1}^{k} d_2^{k-i}\,|e(i)|$$

We can get Equation (31) by combining (3), (9), (10) and (17) together.

$$e(k+1) = y^* - y(k+1) = y^* - y(k) - \boldsymbol{\phi}_L^T(k)\Delta\boldsymbol{U}_L(k)$$
$$= e(k) - \boldsymbol{\phi}_L^T(k)[\boldsymbol{A}(k)\Delta\boldsymbol{U}_L(k-1)$$
$$+\rho_1\boldsymbol{g}^T\left[\hat{\hat{\boldsymbol{\Psi}}}(k)^T\hat{\hat{\boldsymbol{\Psi}}}(k)+\lambda\boldsymbol{I}\right]^{-1}\hat{\hat{\boldsymbol{\Psi}}}^T(k)\boldsymbol{EC}e(k)]$$
$$= (1-\rho_1\phi_1(k)\boldsymbol{g}^T\left[\hat{\hat{\boldsymbol{\Psi}}}^T(k)\hat{\hat{\boldsymbol{\Psi}}}(k)+\boldsymbol{\lambda}\boldsymbol{I}\right]^{-1}\hat{\hat{\boldsymbol{\Psi}}}^T(k)\boldsymbol{E})\bullet \tag{31}$$
$$e(k)+\boldsymbol{\phi}_L^T(k)\boldsymbol{A}(k)\Delta\boldsymbol{U}_L(k-1)$$
$$= (1-\rho_1\phi_1(k)\boldsymbol{P})e(k)+\boldsymbol{\phi}_L^T(k)\boldsymbol{A}(k)\Delta\boldsymbol{U}_L(k-1)$$

Similarly, there exists a positive $\lambda_{\min 3}$ and a positive $M_3$, such that $\lambda > \lambda_{\min 3}$, then we get

$$
\begin{aligned}
0 < M_3 &\le \left|\rho_1\phi_1(k)\boldsymbol{P}\right| \\
&= \left|\rho_1\phi_1(k)\boldsymbol{g}^T\left[\hat{\hat{\boldsymbol{\Psi}}}^T(k)\hat{\hat{\boldsymbol{\Psi}}}(k)+\boldsymbol{\lambda I}\right]^{-1}\hat{\hat{\boldsymbol{\Psi}}}^T(k)\boldsymbol{E}\right| \\
&\le \rho_1\left|\phi_1(k)\right|\sqrt{N_u}\,\frac{1}{\min\limits_{i=1,\cdots Nu}\{\lambda+b_i\}}\left\|\hat{\hat{\boldsymbol{\Psi}}}^T(k)\right\|_\infty < 0.5
\end{aligned} \tag{32}
$$

According to (32), we have

$$
0.5 < \left|1-\rho_1\phi_1(k)\boldsymbol{P}\right| \le \left|1-\rho_1\left\|\phi_1(k)\boldsymbol{P}\right\|\right| \le 1-M_3 < 1 \tag{33}
$$

Let $d_4 = 1-M_3$ and take the norm of (31), then we have

$$
\begin{aligned}
\left|e(k+1)\right| &= \left|1-\rho_1\phi_1(k)\boldsymbol{P}\right|\left|e(k)\right| + \left\|\boldsymbol{\phi}_L^T(k)\right\|_v\left\|\boldsymbol{A}(k)\right\|_v\left\|\Delta\boldsymbol{U}_L(k-1)\right\|_v \\
&< d_4\left|e(k)\right| + d_2\left\|\boldsymbol{\phi}_L^T(k)\right\|_v\left\|\Delta\boldsymbol{U}_L(k-1)\right\|_v \\
&< \cdots < d_4^k\left|e(1)\right| + d_2\sum_{i=1}^{k-1}d_4^{k-1-i}\left\|\boldsymbol{\phi}_L^T(i+1)\right\|_v\left\|\Delta\boldsymbol{U}_L(i)\right\|_v \\
&< d_4^k\left|e(1)\right| + d_2\sum_{i=1}^{k-1}d_4^{k-1-i}\left\|\boldsymbol{\phi}_L^T(i+1)\right\|_v\rho_1 M_2\sum_{j=1}^{i}d_2^{i-j}\left|e(j)\right|
\end{aligned} \tag{34}
$$

Then (34) can be rewritten as

$$
\left|e(k+1)\right| < d_4^k\left|e(1)\right| + d_2 d_3\sum_{i=1}^{k-1}d_4^{k-1-i}\sum_{j=1}^{i}d_2^{i-j}\left|e(j)\right| \tag{35}
$$

Let

$$
g(k+1) = d_4^k\left|e(1)\right| + d_2 d_3\sum_{i=1}^{k-1}d_4^{k-1-i}\sum_{j=1}^{i}d_2^{i-j}\left|e(j)\right| \tag{36}
$$

Then, inequation (35) can be rewritten as

$$
\left|e(k+1)\right| < g(k+1),\ k=1,2,\ldots \tag{37}
$$

where, $g(2) = d_4\left|e(1)\right|$, $d_4 = 1-M_3 > 0.5 > d_3$

Based on (36) and (37), we get

$$
\begin{aligned}
&g(k+2)-g(k+1) \\
&= (d_4-1)d_4^k\left|e(1)\right| + (d_4-1)d_2 d_3\sum_{i=1}^{k-1}d_4^{k-1-i}\sum_{j=1}^{i}d_2^{i-j}\left|e(j)\right| \\
&\quad + d_2 d_3\sum_{j=1}^{k}d_2^{k-j}\left|e(j)\right| \\
&< (d_4-1)g(k+1) + d_2 d_3\sum_{j=1}^{k-1}d_2^{k-j}\left|e(j)\right| + d_2 d_4\left|g(k)\right| \\
&< (d_4-1)g(k+1) + d_2 d_3\sum_{j=1}^{k-1}d_2^{k-j}\left|e(j)\right| + d_2 d_4(d_4^{k-1}\left|e(1)\right| \\
&\quad - d_2 d_3\sum_{i=1}^{k-2}d_4^{k-2-i}\sum_{j=1}^{i}d_2^{i-j}\left|e(j)\right|) \\
&= (d_4-1)g(k+1) + d_2(d_4^k\left|e(1)\right| - d_2 d_3\sum_{i=1}^{k-1}d_4^{k-1-i}\sum_{j=1}^{i}d_2^{i-j}\left|e(j)\right|) \\
&= (d_4+d_2-1)g(k+1)
\end{aligned} \tag{38}
$$

Substituting (38) into (37), we get

$$
g(k+2) < (d_4+d_2)g(k+1) \tag{39}
$$

When $0<\rho_i<1$, $(i=2,\cdots,L+1)$, we have $0 < (\max\limits_{i=2,\cdots,L+1}\rho_i)^{1/Ly+Lu-1}M_1 < M_3 < 1$, then we further get

$$
d_4 + d_2 = 1-M_3+(\max_{i=1,\cdots,Ly+Lu}\rho_i)^{1/Ly+Lu-1}M_1 < 1 \tag{40}
$$

Substituting (40) into (39), we have

$$
\lim_{k\to\infty} g(k+2) < \lim_{k\to\infty}(d_4+d_2)g(k+1) < \cdots < \lim_{k\to\infty}(d_4+d_2)^k g(2) = 0 \tag{41}
$$

Therefore, the conclusion 1) of *Theorem 2* is the result of (41) and (37) when $\lambda > \lambda_{\min} = \max\{\lambda_{\min 1,},\lambda_{\min 2},\lambda_{\min 3}\}$.

Based on (30), (36), (37) and (39), we have

$$
\begin{aligned}
\left\|\boldsymbol{U}_L(k)\right\|_v &\le \sum_{i=0}^{k}\left\|\Delta\boldsymbol{U}_L(i)\right\|_v \\
&\le \sum_{j=1}^{k}\left[d_2^j\left\|\Delta\boldsymbol{U}_L(0)\right\|_v + \rho_1 M_2\sum_{i=1}^{j}d_2^{j-i}\left|e(i)\right|\right] \\
&< \rho_1 M_2\sum_{j=1}^{k}\sum_{i=1}^{j}d_2^{j-i}\left|e(i)\right| \\
&= \rho_1 M_2[(\left|e(1)\right| + (d_2^1\left|e(1)\right| + \left|e(2)\right|) \\
&\quad +(d_2^2\left|e(1)\right| + d_2^1\left|e(2)\right| + \left|e(3)\right|)\cdots + (d_2^{k-1}\left|e(1)\right| \\
&\quad + \cdots + \left|e(k)\right|)] \\
&< \frac{\rho_1 M_2}{1-d_2}(\left|e(1)\right| + \left|e(2)\right| + \cdots + \left|e(k)\right|) \\
&< \frac{\rho_1 M_2}{1-d_2}(\left|e(1)\right| + \left|g(2)\right| + \cdots + \left|g(k)\right|) \\
&< \frac{\rho_1 M_2}{1-d_2}\left(\left|e(1)\right| + \frac{\left|g(2)\right|}{1-d_2-d_4}\right)
\end{aligned} \tag{42}
$$

Hence, (42) proves the boundedness of $\left\|\boldsymbol{U}_L(k)\right\|_v$. The conclusion 2) of Theorem 2 in Section III is proved.

We finished the proof of *Theorem* 2.